\documentclass[twocolumn,showpacs,amsmath,amssymb,groupedaddress]{revtex4}

\usepackage{graphicx}
\usepackage[dvipdfm]{hyperref}

\begin{document}

\title{Scanning magnetoresistance microscopy of atom chips}

\author{M. Volk}
%\email[Electronic address: ]{MVolk@swin.edu.au}
\author{S. Whitlock}
\altaffiliation[Present address: ]{Van der Waals-Zeeman Instituut,
Universiteit van Amsterdam,
Valckenierstraat 65-67,
1018 XE Amsterdam,
The Netherlands
}
\author{B. V. Hall}
\email[Electronic address: ]{BrHall@swin.edu.au}
\author{A. I. Sidorov}
\affiliation{ARC Centre of Excellence for Quantum-Atom Optics and\\
Centre for Atom Optics and Ultrafast Spectroscopy,\\
Swinburne University of Technology, Hawthorn, Victoria 3122,
Australia}

\date{\today}

\begin{abstract} Surface based geometries of microfabricated wires
or patterned magnetic films can be used to magnetically trap and
manipulate ultracold neutral atoms or Bose-Einstein condensates.  We
investigate the magnetic properties of such atom chips using a
scanning magnetoresistive (MR) microscope with high spatial
resolution and high field sensitivity.  We show that MR sensors are
ideally suited to observe small variations of the magnetic
field caused by imperfections in the wires or magnetic materials
which ultimately lead to fragmentation of ultracold atom clouds.
Measurements are also provided for the magnetic field produced by a
thin current-carrying wire with small geometric modulations along
the edge.  Comparisons of our measurements with a full numeric
calculation of the current flow in the wire and the subsequent
magnetic field show excellent agreement. Our results highlight the
use of scanning MR microscopy as a convenient and powerful technique
for precisely characterizing the magnetic fields produced near the
surface of atom chips. \end{abstract}


\maketitle

\section{Introduction}

Surface based potentials for manipulating neutral atoms on a
micron scale have attracted widespread interest in recent years.
Atom chips~\cite{Fol02,Fort07} consisting of planar geometries of
microfabricated wires or patterned magnetic materials provide
intricate magnetic potentials and have become a practical and robust
tool for producing, trapping and manipulating Bose-Einstein
condensates.  Atoms chips have recently been used to precisely
position Bose-Einstein condensates~\cite{gun05}, realize trapped
atom interferometers~\cite{sch05,jo07} and have provided new and
sensitive techniques for detecting tiny forces on a small spatial
scale~\cite{hal06b}.  The fabricated wires or magnetic materials
used for atom chips have been the topic of several recent studies,
finding that their quality must be exceptionally high since even the
smallest imperfections, for example roughness of the wire edge, can
lead to uncontrolled magnetic field variations. These variations
subsequently corrugate the bottom of the trapping
potential~\cite{wan04,est04}.  Recently, fragmentation of ultracold
atoms has also been observed in close proximity to magnetic
materials~\cite{sin05a,whit07,boy06} and has been traced to long
range spatial variations in the film magnetization~\cite{whit07}. As
the energy scales associated with ultracold atoms and Bose-Einstein
condensates are in the nanokelvin regime, even the smallest magnetic
field variations of only a few nanotesla can dramatically alter
their properties~\cite{wil05}.

Until now, characterizing the smoothness of the potentials produced
by atom chips has relied on the atom clouds themselves, through
either the equilibrium atomic density distribution~\cite{est04} or
radio frequency spectroscopy of trapped atom clouds~\cite{whit07}.
With the increasing complexity of atom chips, however, it is
necessary to obtain {\it fast} and {\it reliable} methods of
characterizing the magnetic potentials prior to installing the atom
chips in ultrahigh vacuum and trapping ultracold atoms.

\begin{figure}
    \includegraphics[width=0.9\columnwidth]{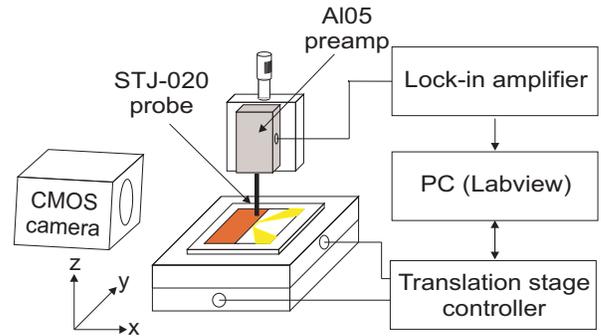}
    \caption{(color online) Schematic of the scanning magnetoresistance
    microscope. The sample is placed on a computer controlled x-y
    translation stage. The magnetoresistive probe is connected to a
    preamplifier and the signal is filtered and digitized by a lock-in
    amplifier. A CMOS camera is used to determine the distance
    between the sensor tip and the sample.
    \label{Setup}}
\end{figure}

In this paper we describe the application of a micron sized
magnetoresistance (MR) sensor to accurately profile the magnetic
fields generated by magnetic film and current-carrying wire atom
chips.  Our home-built magnetoresistance microscope
(Fig.~\ref{Setup}) is used to measure small magnetic field
variations above a permanent magnetic film atom chip which causes
fragmentation of ultracold atom clouds.  The MR measurements support
independent measurements performed using trapped ultracold atoms as
the magnetic field probe~\cite{whit07}.  The study indicates the
variations occur predominately near the edge of the film and are
associated with heating of the film during vacuum bake-out.  In
addition, we have fabricated a new current-carrying wire atom chip
using femtosecond laser ablation of a gold film~\cite{wolf07}. A
wire is sculpted with a periodically modulated edge to produce a
complex magnetic potential for ultracold atoms.  Two dimensional
images of the field produced by the wire are obtained and are in
excellent agreement with numeric calculations of the expected field
strength. The measurements show that it is possible to fabricate and
characterize a linear array of magnetic potentials produced by
modifying the edge of a straight current carrying wire.

\section{Apparatus}

Analysis of the atom chips is performed using an ultra-sensitive
low-field magnetoresistive sensor based on magnetic tunneling
junction technology~\cite{moo95,lac02}.  A magnetic tunneling
junction sensor consists of two ferromagnetic layers separated by an
ultra-thin insulating interlayer.  One magnetic layer has fixed
`pinned' magnetization while the other responds to the local
magnetic field.  The interlayer resistance depends on the relative
magnetization of the neighboring magnetic layers. These devices provide an
absolute measure of the magnetic field with high sensitivity and
high spatial resolution. They provide a linear response over a large
field range (typically about 0.5 mT) and are ideal for studying the
magnetic fields produced by microfabricated current-carrying wires
or patterned magnetic materials on atom chips.  Here the sensor is
incorporated into a home-built scanning magnetic field microscope,
schematically depicted in Figure~\ref{Setup}, and used to study the
corrugated field produced by the atom chip.

The microscope (Fig.~\ref{Setup}) consists of the MR sensor probe,
the preamplification electronics, a lock-in amplifier, a motorized
x-y translation stage and a computer interfaced via LabView to both
stage and lock-in amplifier. The probe tip is manually positioned
above the sample using a micrometer stage and a CMOS camera for
height calibration. This setup allows us to acquire one-dimensional
scans as well as two-dimensional maps of the $z$-component, i.e. the
out-of-plane component, of the magnetic field at variable heights
above the sample surface.

Our scanning magnetoresistance microscope incorporates a
commercially available magnetic tunnel junction probe
(MicroMagnetics STJ-020), polished to allow very
close approaches to the surface ($\sim\,$10~$\mu$m).  The active
area of the sensor is approximately $5\times5~\mu$m$^2$ and it
detects the magnetic field oriented along the sensor tip ($z$
direction). The sensor is interfaced using an Anderson
loop~\cite{and92} to convert small changes in the sensor resistance
to a signal voltage. The output is then amplified using a signal
amplification board (MicroMagnetics AL-05) with
a gain of 2500 and a bandwidth of 1~MHz. The sensor and preamplifier
are calibrated to give an output of 20~V/mT. Due to its small size
the sensor exhibits significant $1/f$ noise which can be overcome by
reducing the bandwidth of the output signal.

To increase the signal-to-noise ratio we use an AC modulation
technique. In the case of current-carrying wires this is simply done
by modulating the wire current at kHz frequencies and detecting the
signal with a lock-in amplifier (Stanford Research Systems SR830).
When studying permanent magnetic films we use mechanical modulation
of the probe. The tip of the probe is oscillated along the scanning
direction at its mechanical resonance frequency (18~kHz) using a
piezo actuator. At this frequency the noise level of the sensor is
reduced to less than 15\% compared to DC; however the output of
the lock-in amplifier is now proportional to the first derivative of
the magnetic field. This output is calibrated against a known
magnetic field gradient by first measuring the field in DC mode
200~$\mu$m above the edge of the film. The field is large enough to
provide good signal-to-noise and features a large gradient of
1~Tm$^{-1}$. We then compare the numerical derivative of this
measurement to the data obtained while oscillating the tip. This
allows us to determine the oscillation amplitude of the probe and
hence to reconstruct the magnetic field up to a constant offset by
numerical integration of the data. The oscillation amplitude and
subsequently the spatial resolution of this measurement is
approximately 50~$\mu$m. The AC modulation technique reduces the
noise levels to about $0.1~\mu$T, equivalent to that obtained using
ultracold atoms as a probe \cite{whit07} and a factor of 5 lower
than what is obtained for an equivalent measurement time using just
low-pass filtering.

\section{Permanent magnet atom chip}
\label{Section_perm_magn_atom_chips}

As a first application of the magnetic field microscope we
investigated the random variations in the magnetic potential created near the
surface of a magnetic film atom chip used in previous experiments to trap
ultracold atoms and Bose-Einstein condensates and is described in
detail elsewhere \cite{hal06}. It uses a multilayer
Tb$_6$Gd$_{10}$Fe$_{80}$Co$_4$ film which exhibits strong
perpendicular anisotropy. The film is deposited on a 300~$\mu$m
thick glass substrate where one edge is polished to optical quality
prior to film deposition. At this edge the magnetic film produces a
field that is analogous to that of a thin current-carrying wire
aligned with the edge (I$_{\mbox{eff}}$ = 0.2 A). A magnetic
microtrap is formed by the field from the film, a uniform magnetic
bias field, and two current-carrying end-wires. To account for the
need of a reflecting surface for the mirror magneto-optical trap the
chip is completed by a second glass slide and both sides are coated
with gold.

\begin{figure}
    \includegraphics[width=\columnwidth]{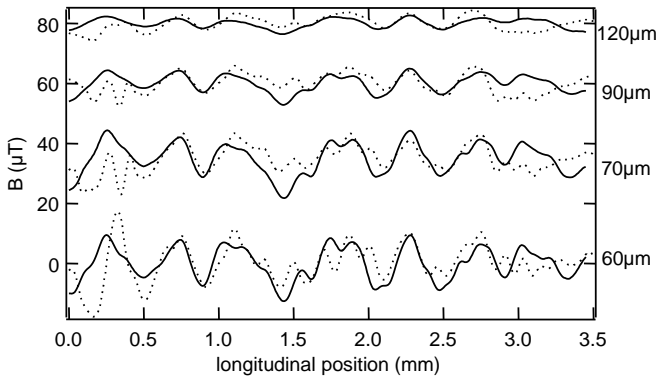}
    \caption{Magnetic field profiles at various distances
    above the magnetic film edge of a permanent magnetic atom chip
    measured with the magnetoresistance
    microscope sensitive to the $B_z$ field component (solid lines).
    The dotted lines correspond to measurements of the magnetic field at approximately the same distance using an ultracold
    atom cloud sensitive to the $B_y$ field component. The profiles have been offset for clarity.
    The relative longitudinal offset between the two measurements
    is initially unknown and is adjusted for optimum agreement.
    \label{B_xy_Atoms_and_MR}}
\end{figure}

Due to their narrow energy distribution, ultracold atoms are very
sensitive to small fluctuations of the magnetic trapping potential.
In a recent paper \cite{whit07} we used radio frequency (rf)
spectroscopy of trapped atoms to measure the absolute magnetic field
strength above the edge of the film. This provided an accurate
measurement of the corrugation of the longitudinal component of the
magnetic field produced by the permanent magnetic atom chip, i.e.
the component parallel to the film edge. We also developed a model
describing the spatial decay of random magnetic fields from the
surface due to inhomogeneity in the film magnetization.

After removing the atom chip from the vacuum chamber we used the
magnetoresistance microscope to further characterize the film
properties. Our first measurement consists of a series of scans of
the magnetic field parallel to the film edge over a region of 3.5~mm
at various heights ranging from $500~\mu$m down to $60~\mu$m, the
minimum distance limited by the adjacent protruding gold coated
glass slide. Four of these profiles are depicted in
Figure~\ref{B_xy_Atoms_and_MR}. Due to the large field gradient
produced at the film edge it was necessary to carefully align the
measurement direction and subtract a third order polynomial from the
data. Also plotted in the same figure are the corresponding profiles
previously measured by rf spectroscopy of ultracold atoms.  The results from the two
different methods are in remarkable agreement.  It should be noted
however that a quantitative comparison is difficult as the two
methods are sensitive to different components of the corrugated
magnetic field: the magnetoresistive sensing direction is
perpendicular to the surface while the trap bottom probed by the rf
spectroscopy is defined by the in-plane component of the magnetic
field.

\begin{figure}
    \includegraphics[width=\columnwidth]{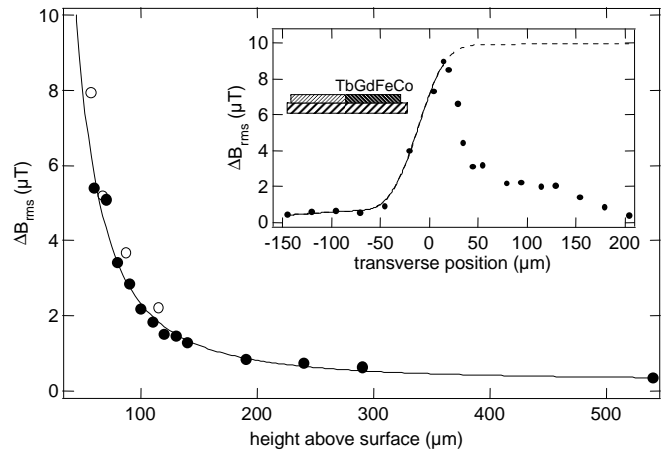}
    \caption{Behavior of the magnetic field roughness
    $\Delta B_{rms}$ above the film edge
    measured using the magnetoresistance microscope (filled circles) and
    rf spectroscopy of ultracold atoms (open circles), as a function of distance from the film surface.
    The solid line is a power-law fit to the magnetoresistance microscope data.
    The inset shows the dependence of the field roughness on the
    transverse distance from the film edge for a fixed height of $z_0 = 60~\mu$m above the film
    surface.
    \label{DeltaB_vs_height}}
\end{figure}

The results of the complete series of magnetoresistance scans as
well as the rf spectroscopy measurements are summarized in
Figure~\ref{DeltaB_vs_height} where the root mean square ({\it rms})
noise is plotted as a function of distance to the surface. For
random white noise fluctuations of the film magnetization our model
described in \cite{whit07} predicts a $z^{-2}$ decay of the field
roughness. A power law fit to the data obtained by the MR scans
gives $\Delta B_{rms} \propto z^{-1.9 \pm 0.2}$ in excellent
agreement with this prediction.

We have also performed a series of scans at constant height ($z_0 =
60~\mu$m) above the film surface but variable transverse distance to
the film edge. The {\it rms} noise levels of these scans are
depicted in the inset of Figure~\ref{DeltaB_vs_height} (circles)
together with the prediction of the random magnetization model
(lines). While the model describes the results adequately above the
non-magnetic half plane of the atom chip the measured inhomogeneity
decreases away from the edge above the magnetic film side, whereas
in the case of homogeneous magnetization fluctuations $\Delta B_{rms}$
is expected to stay constant (dotted line in
Fig.~\ref{DeltaB_vs_height}).

\section{Tailored magnetic microtraps}

Section~\ref{Section_perm_magn_atom_chips}  of this paper focused on
MR studies of the corrugated potential produced by a partially
inhomogeneous magnetic film atom chip.
%Our magnetoresistance microscope is also suitable for
%characterising the magnetic potentials produced by microfabricated
%current-carrying wire atom chips.
In this section, we describe the analysis of a current-carrying wire
atom chip fabricated using micron-scale femtosecond laser ablation
of a thin metal film.  We have produced a tailored magnetic
potential by sculpting the shape of a wire to create a linear array
of magnetic traps for cold atoms.  Two-dimensional magnetoresistance
microscopy provides an image of the perpendicular magnetic field
component produced by the wire at a fixed distance to the surface. A
solution to the magnetostatic inverse problem is then applied to
obtain the remaining two field components, allowing a complete
reconstruction of the magnetic trapping potential.  Of particular
interest is the field component parallel to the wire, which defines
the bottom of the trapping potential.  A comparison of the measured
and reconstructed field components with full numeric calculations of
the field produced by the sculptured wire shows excellent agreement.

\subsection{Sculptured wire atom chip}

Femtosecond laser ablation can be used to pattern micron and
submicron scale structures on a wide variety of materials
\cite{Nol97} and can be used to produce atom chips~\cite{wolf07}.
%In particular, femtosecond laser ablation of metal films is receiving
%attention as a means to rapidly produce photomasks for subsequent
%lithographic processes in a cost effective way [ref,ref].
In this work we use the technique to directly fabricate complex wire
patterns in an evaporatively deposited gold film to form a
current-carrying wire atom chip~\cite{Fort07}.  The chip consists of
a glass slide substrate with a 25~nm thick Cr bonding layer and a
150~nm thick Au layer. The wire structure is patterned by cutting
three $3~\mu$m wide insulating channels into the Au film.

\begin{figure}
    \includegraphics[width=0.9\columnwidth]{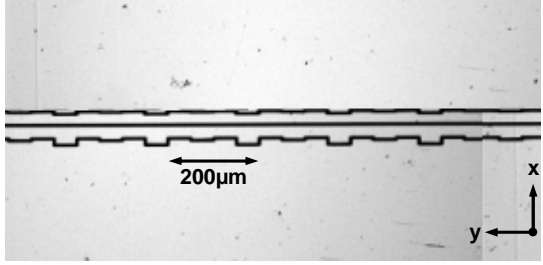}
    \caption{Optical microscope image of the current-carrying wire atom chip.
    The two sculptured wires are formed by cutting three $3~\mu$m wide insulating
    channels, visible as black lines,
    into a 150~nm thick Au layer using fs laser ablation. \label{WireSample}}
\end{figure}

We have patterned two parallel Au wires with widths of 20 and 30
$\mu$m and lengths of 10~mm which can be used to create a magnetic
potential for trapping Bose-Einstein condensates
(Fig.~\ref{WireSample}). Each wire has been sculptured with one
periodically modulated boundary with a period of $200~\mu$m.
Deliberately modulating the wire boundary slightly modifies the
current path and produces a small field component oriented parallel
to the wire, which modulates the corresponding longitudinal magnetic
potential experienced by the trapped atoms \cite{Piet05}. This is
used to realize a linear array of asymmetric double wells which are
separated by potential barriers with small amplitudes which can be
precisely controlled by varying the wire current or the distance of
the trap to the wire surface~\cite{sch05,hal06b,sid06}.
%This sculptured wire method allows versatile magnetic potentials to be
%produced for manipulating Bose-Einstein condensates.

The $30~\mu$m wire is chosen for the magnetoresistance measurements.
We use the reference source of the lock-in amplifier to drive a
small AC current of 37~mA {\it rms} through the wire at a frequency
of 1~kHz.  The output of the lock-in amplifier is recorded by a
computer.  Two computer controlled translations stages are used to
position the wire sample with respect to the MR probe. The probe is
calibrated against the expected field produced by the wire
calculated using Biot-Savart's law, neglecting the effect of the
small modulations. We record an image of the perpendicular magnetic
field component produced by the wire over a 2$\times$1.5~mm$^2$
spatial region at a distance of $z=30~\mu$m above the wire.  The
spatial resolution is $10~\mu$m which corresponds to $150\times200$
data points.  The lock-in integration time is set to 300~ms and each
line of the image is scanned twice and averaged, which results in a
measurement time of approximately 5 hours for the whole
two-dimensional magnetic field image. Figure~\ref{TwoD_plots}a shows
the result of this measurement (only the central part of the full
image is shown). The field amplitude produced across the wire at
this height is $\pm100$~$\mu$T. Directly above the wire the
perpendicular field is nearly zero apart from a small modulated
field component with amplitude of about $\pm2.5~\mu$T. The noise
level for this measurement determined from a region about 0.7~mm
away from the wire was as low as 50~nT.

\subsection{Reconstruction of the in-plane field components}

\begin{figure}
    \includegraphics[width=\columnwidth]{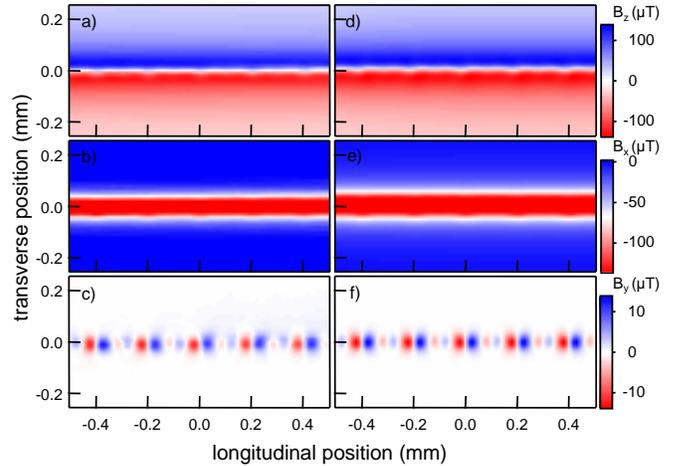}
    \caption{(Color online) a--c: measured out-of-plane component $B_z$ and
    reconstructed in-plane components $B_x, B_y$ of the magnetic field
    above the current-carrying wire atom chip. d--f: corresponding results
    of the numerical simulation of the current distribution and the associated
    magnetic field, based on the geometric dimensions of the wire structure.\label{TwoD_plots}}
\end{figure}

With a two-dimensional image of the out-of-plane field component at
a given height it is possible to convert to a uniquely defined
in-plane current distribution \cite{roth89} and subsequently back to
any other field component.  Given that the height of the wire is small
compared to the measurement distance above the surface, the current
density can be considered as a two-dimensional distribution. The
Fourier transforms of the magnetic field components $b_x$ and $b_y$
are then simply related to $b_z$:

\begin{eqnarray}
b_x(k_x,k_y)=i \frac{k_x}{k} b_z(k_x,k_y)\nonumber \\ \nonumber \\
b_y(k_x,k_y)=i \frac{k_y}{k} b_z(k_x,k_y)
\end{eqnarray}
where $k=\sqrt{k_x^2+k_y^2}$.

Shown in Figure~\ref{TwoD_plots} (b) and (c) are the reconstructed
in-plane field components $B_x$ and $B_y$ derived from the measured
$B_z$ component. The $B_y$ field image clearly shows the modulated
component along the length of the wire which defines the bottom of
the trapping potential.

\subsection{Numeric calculations of magnetic fields}

\begin{figure}
    \includegraphics[width=\columnwidth]{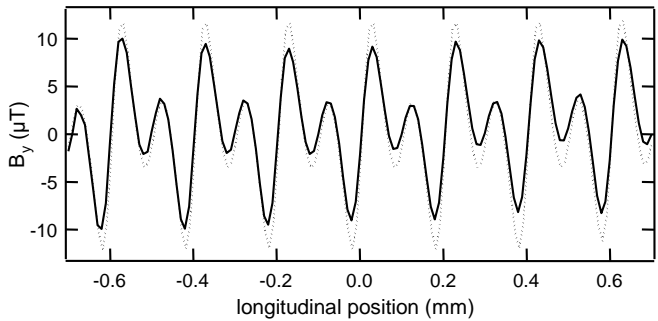}
    \caption{Line profile of the magnetic field component parallel to the wire ($B_y$)
    at $x=0$, i.e. directly above the wire. The
    solid line represents the field data reconstructed from the MR
    measurement while the dotted line shows the simulated values.
    \label{B_parallel}}
\end{figure}

We have also performed detailed numeric calculations of the field
produced by the sculptured wire to compare them with our
measurements. The current density distribution of the wire is
computed from the solution of Laplace's equation
$\nabla\cdot\nobreak(\sigma\nabla\nobreak V)=0$ satisfied by the
electrostatic potential $V$.  We assume that the conductivity
$\sigma$ is uniform throughout the wire and, since we are interested
in the field at distances much larger than the wire thickness, we
assume that $V$ depends only on $x$ and $y$.  Exact analytical
solutions for this problem can be obtained for particular
geometries; however in general one has to rely on numerical methods.

Here, solutions of Laplace's equation were computed using the finite
element method which provides an approximate solution of partial
differential equations with defined boundary conditions.  For this
problem we have used the Matlab Partial Differential Equation (PDE)
toolbox. The boundary conditions are specified such that the normal
component of the current density on the wire edge is zero (Neumann
conditions). The wire geometry is then decomposed into a set of
triangular elements which define a mesh of nodes, for each of which
the electrostatic potential is solved. From this it is straight
forward to compute the current distribution and the associated
magnetic field. The results of these calculations are depicted in
Figure~\ref{TwoD_plots} (d) to (f) next to the corresponding
measurements.

The $y$ component of the magnetic field produced by the atom chip is
of particular interest because it determines the potential minimum
experienced by the trapped atoms. Figure~\ref{B_parallel} compares
field profiles of $B_y$ along the wire extracted from the
measurement and the simulation. We note that the two profiles differ
by about 10\% in amplitude and attribute this to a systematic error
in the calibration of the sensor which was done assuming that the
measured $B_z$ profile was produced by an infinitely thin wire. In
addition to that, the measured field amplitude decreases slightly
over the 2~mm scan region which is most likely due to a tilt between
the sample surface and the measurement plane on the order of 2~mrad.

\section{Conclusion}

We have demonstrated a scanning magnetic microscopy technique for
characterizing atom chips. The microscope is based on a commercially
available magnetoresistive probe. It has been used to scan the
corrugation of the magnetic field produced by a permanent magnet
atom chip as well as to investigate the field produced by a
sculptured current-carrying wire.

The spatial resolution of the device is in principle limited by the
size of the active area of the probe, i.e., about $5~\mu$m for the
sensor used in this work; however submicron resolution has been
demonstrated in similar applications~\cite{sch03}. For our
demonstration the smallest measurable feature sizes were determined
by the minimum distance to the surface ($\geq 10~\mu$m) and the
scaling laws for magnetic fields. The scan range is limited only by
the computer controlled translation stages which can easily be
extended to several centimeters.

By simple low pass filtering of the output signal and averaging we
were able to achieve a sensitivity of 0.5~$\mu$T when measuring a
permanent magnetic film. Using AC modulation techniques we could
reduce this down to 0.1~$\mu$T in case of stationary magnetic fields
and even 50~nT for current carrying wires.

In conclusion, the high field sensitivity, large scan range, ease of
use and low cost makes the magnetoresistance microscope the
quintessential tool for ex-situ characterization of cold atom
magnetic microtraps.

\begin{acknowledgments}

The authors would like to thank J. Wang for the deposition of the
films. This project is supported by the ARC Centre of Excellence for
Quantum-Atom Optics.

\end{acknowledgments}

\bibliography{MRM-of-Atom-Chips}

\end{document}